\documentclass[aps,prl,twocolumn,10pt,superscriptaddress]{revtex4-2}
\usepackage{graphicx}
\usepackage{amsmath,amsfonts}
\usepackage{color}
\usepackage{ulem}
\usepackage{braket}
\usepackage{soul}
\newcommand{\upsub}[1]{\sb{\mathrm{#1}}}
\newcommand{\upsup}[1]{\sp{\mathrm{#1}}}

\begingroup\lccode`~=`_\lowercase{\endgroup\let~\upsub}
\begingroup\lccode`~=`^\lowercase{\endgroup\let~\upsup}

\AtBeginDocument{%
	\catcode`_=12 \catcode`^=12
	\mathcode`_="8000
	\mathcode`^="8000
}

\begin{document}


	\title{Ultrafast dynamics of coherent exciton-polaritons in van der Waals semiconductor metasurfaces}
	
	\author{Luca Sortino}
    \thanks{These authors contributed equally to this work.}
    \affiliation{Chair in Hybrid Nanosystems, Nanoinstitute Munich, Faculty of Physics, Ludwig-Maximilians-Universit{\"a}t M{\"u}nchen, 80539 Munich, Germany}
    \author{Armando Genco}
    \thanks{These authors contributed equally to this work.}
 	\affiliation{Dipartimento di Fisica, Politecnico di Milano, Piazza Leonardo Da Vinci 32, 20133 Milano, Italy}
  	\author{Cristina Cruciano}
    \author{Michele Guizzardi}
    \affiliation{Dipartimento di Fisica, Politecnico di Milano, Piazza Leonardo Da Vinci 32, 20133 Milano, Italy}
    \author{Daniel Timmer}
	\affiliation{Physics Department and Center for Nanoscale Dynamics (CeNaD), Carl von Ossietzky Universität Oldenburg, D-26129 Oldenburg, Germany}
    \author{Thomas Weber}
	\affiliation{Chair in Hybrid Nanosystems, Nanoinstitute Munich, Faculty of Physics, Ludwig-Maximilians-Universit{\"a}t M{\"u}nchen, 80539 Munich, Germany}
    \author{Jonathan O. Tollerud}
    \affiliation{Optical Sciences Centre, Swinburne University of Technology, Hawthorn, Victoria 3122, Australia}
    \affiliation{ARC Centre of Excellence in Future Low-Energy Electronics Technologies, Swinburne University of Technology, Hawthorn, Victoria 3122, Australia}
    \author{Francesco Gucci}
    \author{Matteo Corti}
    \author{Gianluca Valentini}
	\affiliation{Dipartimento di Fisica, Politecnico di Milano, Piazza Leonardo Da Vinci 32, 20133 Milano, Italy}
  	\author{Cristian Manzoni}
    \affiliation{CNR-IFN, Piazza Leonardo da Vinci 32, Milano, 20133, Italy}
	\affiliation{Dipartimento di Fisica, Politecnico di Milano, Piazza Leonardo Da Vinci 32, 20133 Milano, Italy}
    \author{Stefano Dal Conte}
	\affiliation{Dipartimento di Fisica, Politecnico di Milano, Piazza Leonardo Da Vinci 32, 20133 Milano, Italy}
    \author{Christoph Lienau}
	\affiliation{Physics Department and Center for Nanoscale Dynamics (CeNaD), Carl von Ossietzky Universität Oldenburg, D-26129 Oldenburg, Germany}
    \author{Jeffrey A. Davis}
    \affiliation{Optical Sciences Centre, Swinburne University of Technology, Hawthorn, Victoria 3122, Australia}
    \affiliation{ARC Centre of Excellence in Future Low-Energy Electronics Technologies, Swinburne University of Technology, Hawthorn, Victoria 3122, Australia}
	\author{Stefan A. Maier}
	\affiliation{School of Physics and Astronomy, Monash University, Clayton, Victoria 3800, Australia}
	\affiliation{The Blackett Laboratory, Department of Physics, Imperial College London, London, SW7 2BW, United Kingdom}
	\author{Andreas Tittl}
     \email{andreas.tittl@physik.uni-muenchen.de}
	\affiliation{Chair in Hybrid Nanosystems, Nanoinstitute Munich, Faculty of Physics, Ludwig-Maximilians-Universit{\"a}t M{\"u}nchen, 80539 Munich, Germany}
 	\author{Giulio Cerullo}
    \email{giulio.cerullo@polimi.it}
	\affiliation{Dipartimento di Fisica, Politecnico di Milano, Piazza Leonardo Da Vinci 32, 20133 Milano, Italy}
    \affiliation{CNR-IFN, Piazza Leonardo da Vinci 32, Milano, 20133, Italy}
	\date{\today}
	\maketitle

\newpage

\noindent
\textbf{Enabling coherent light-matter interactions is a critical step toward next-generation quantum technologies. However, achieving this under ambient temperature conditions remains challenging due to rapid dephasing in optically excited systems. Optical metasurfaces based on quasi-bound states in the continuum have recently emerged as a powerful platform for reaching the strong light-matter coupling regime in flat, subwavelength thickness devices. Here, we investigate ultrafast exciton-polariton dynamics in self-hybridized WS$_2$ thin-film metasurfaces. Using hyperspectral momentum-resolved imaging, we reconstruct the highly anisotropic exciton-polariton dispersion, with a transition from positive to negative effective mass along orthogonal symmetry axes. 
Femtosecond pump-probe and multidimensional spectroscopy reveal detuning-dependent polariton dynamics with a coherence time up to $\sim$110 fs, and allow direct observation of the coherent dynamics through ultrafast Rabi oscillations with $\sim$45 fs period. We describe this behaviour with a three-eigenstate model that couples the photonic resonance with both bright and dark excitons, extending the conventional two-state picture of strong coupling.
Our results establish van der Waals metasurfaces as a promising platform for next-generation polaritonic devices, enabling coherent quantum transfer of matter excitations at room temperature.
}

\bigskip
\noindent
Excitons, tightly Coulomb-bound electron-hole pairs, play a central role in defining the optical properties of semiconductors \cite{hopfield1958theory}. 
When embedded in optical cavities, they can interact coherently with confined photons, leading to the formation of exciton-polaritons, hybrid quasiparticles that inherit both the light effective mass of photons and the strong nonlinearity of matter excitations \cite{kavokin2017microcavities}.
These polaritons emerge in the strong-coupling regime, where the coherent energy exchange rate between excitons and confined photons surpasses their respective dissipation rates. In this regime, the hybrid light-matter states manifest as an anticrossing in the energy spectrum, the so-called normal mode splitting (Rabi splitting) \cite{thompson1992observation}, equal to $\hbar\Omega_R=2g$, where $g$ is the coupling strength.
Such coherent interaction also leads to Rabi oscillations, a periodic exchange of energy between an ensemble of strongly coupled excitons and the cavity mode, with a timescale set by the coupling rate \cite{norris1994time}.
Owing to their bosonic nature, polaritons can collectively occupy the same quantum state, allowing the spontaneous formation of polariton condensates \cite{kasprzak2006bose,deng2003polariton,AmoNat2009} even under non-resonant conditions \cite{byrnes2014exciton}. Furthermore, the high nonlinearities make them promising candidates for next-generation photonic devices, from coherent light sources to quantum simulators \cite{zasedatelev2021single,kavokin2022polariton,liew2023future,genco2025femtosecond}.

Traditionally, achieving strong light-matter coupling has relied on mirror-based microcavities, reaching long polariton coherence times exploiting high quality (Q) factor photonic modes, but with large mode volumes and usually requiring cryogenic operation \cite{kavokin2017microcavities}.
Plasmonic nanostructures can shrink the mode volumes below the diffraction limit of light, allowing the observation of exciton-plasmon polaritons \cite{torma2014strong}, although generally affected by non-radiative losses \cite{wang2014interplay} unless exploiting sub-radiant damping \cite{ropers2005femtosecond}.
Recently, optical metasurfaces have emerged as a transformative platform for enhancing light-matter interactions on a sub-wavelength scale \cite{guan2022light}. Such systems consist of two-dimensional (2D) arrays of metallic or dielectric nanoresonators tailored to host optical resonances at specific frequencies, enabling precise control over the amplitude, phase, and polarization of light \cite{yu2011light,kuznetsov2024roadmap}. 

By leveraging the physics of bound states in the continuum (BICs) \cite{hsu2016bound,koshelev2021}, optical metasurfaces can be engineered to sustain resonances with extremely high Q factors and enable enhanced light-matter interactions as in conventional cavities \cite{Liu2019d,biechteler2025fabrication}. 
BIC are non-radiative (dark) modes arising within the radiation continuum via carefully tailored mode interference \cite{friedrich1985interfering}. They have been transformative in photonics because of their exceptional light trapping capabilities and intrinsic topological nature \cite{zhou2025bound}, enabling applications in nonlinear optics, sensing and optical signal manipulation \cite{kang2023applications}.
However, for efficient light-matter coupling, the dark BIC mode has to be transformed into a bright quasi-BIC (qBIC) resonance \cite{overvig2020selection}. In the case of dielectric metasurfaces, this is achieved by breaking the in-plane inversion symmetry of the unit cell geometry via rational design choices \cite{koshelev2018asymmetric}. In this way, a symmetry-protected radiative channel can be engineered within the photonic environment, realizing tunable high Q factor resonances controlling both the spectral position and the linewidth, allowing precise control of electromagnetic field enhancement.
\begin{figure*}
	\centering
	\includegraphics[width=1\linewidth]{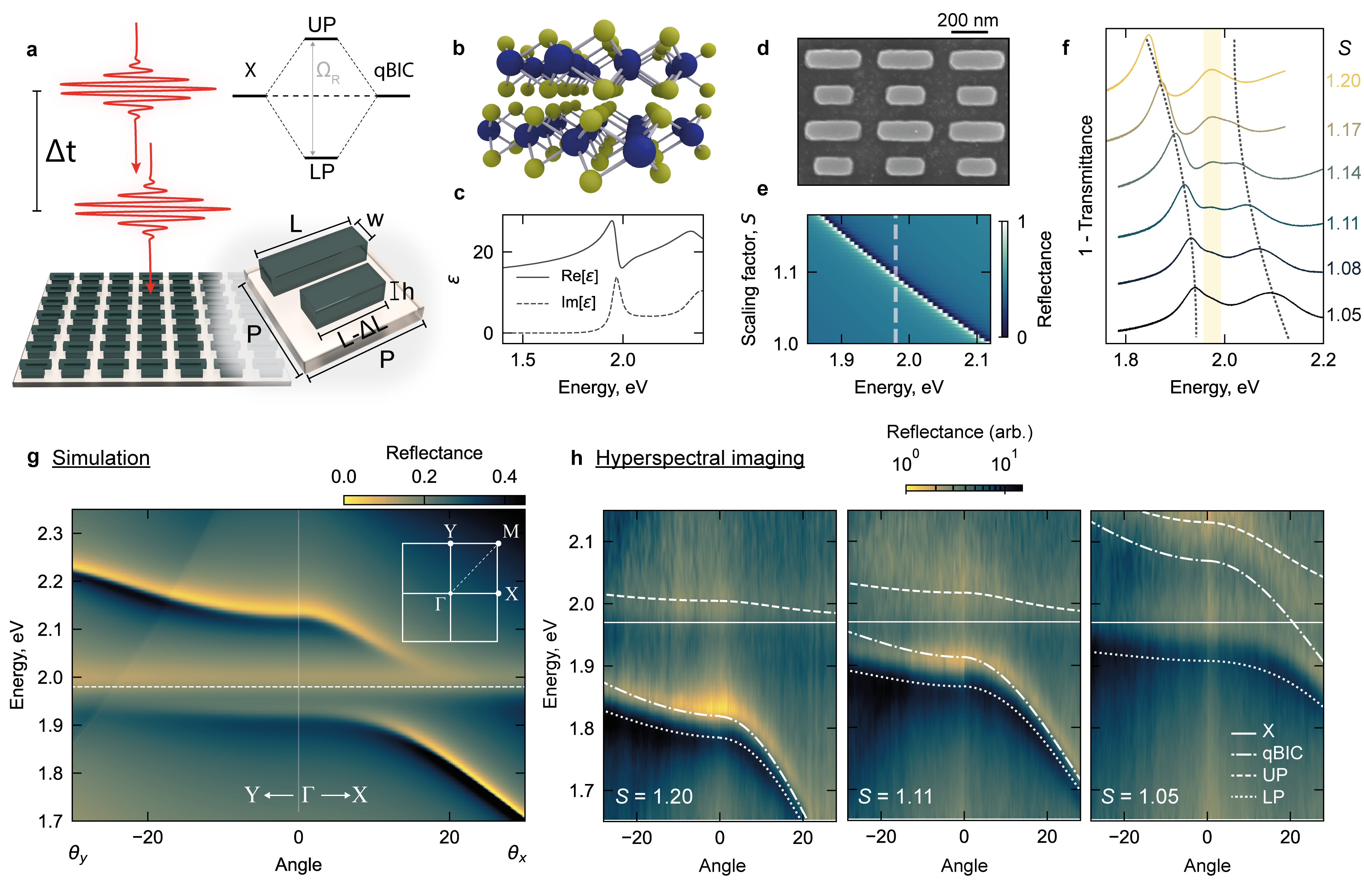}
\caption{\textbf{Photonic band structure of exciton-polaritons in TMDC qBIC metasurfaces}
(a) Schematic of the pump-probe spectroscopy experiment on strongly coupled qBIC metasurfaces, where $\Delta t$ denotes the time delay between pump and probe laser pulses. Bottom inset: Illustration of a single metasurface unit cell and its key design parameters, defined, for a scaling factor $S=1$, by a periodicity of $P_{0} = 340~\mathrm{nm}$, a base rod length of $L_{0} = 266~\mathrm{nm}$, and a rod width of w$_{0} = 90~\mathrm{nm}$. Top inset: Energy level diagram showing the strong coupling between excitons (X) and the qBIC mode, and the upper (UP) and lower (LP) polariton branches separated by the Rabi frequency ($\Omega_R$).
(b) Crystal structure of a 2H-phase TMDC. In yellow the chalcogen atoms (S), in blue the transition metal ones (W).
(c) Real and imaginary parts of the dielectric function of WS$_2$. Adapted from Ref.~\cite{munkhbat2023nanostructured}.
(d) Scanning electron micrograph of a WS$_2$ metasurface fabricated on a glass substrate.
(e) Numerically calculated qBIC resonance energy as a function of the scaling factor $S$ ($\Delta L = 25$ nm). The material is modelled as homogeneous with refractive index $n = 4$, approximating WS$_2$ while neglecting excitonic effects.
(f) Normal-incidence reflectance spectra for WS$_2$ metasurfaces with different $S$ values. The shaded area indicates the spectral position of the exciton. Dashed lines depict the polariton dispersion.
(g) Rigorous Coupled-Wave Analysis (RCWA) simulation of momentum-resolved normalized reflectance for a strongly coupled WS$_2$ metasurface ($\Delta L_0 = 75$ nm), in vacuum. The data are plotted along the $\Gamma \rightarrow X$ direction ($\theta_x$) for positive incidence angles and along $\Gamma \rightarrow Y$ ($\theta_y$) for negative angles. Inset: Brillouin zone of the qBIC metasurface.
(h) Experimental hyperspectral angle-resolved reflectance imaging of fabricated WS$_2$ metasurfaces on glass, shown for different $S$ values.}
\label{fig:fig1}
\end{figure*}

While metasurfaces have predominantly employed high-index dielectrics or metals, the emerging class of van der Waals (vdW) materials opens new avenues for nanophotonics by exploiting their layered structure, intrinsic anisotropy, and ease of vertical stacking \cite{ling2021all,munkhbat2023nanostructured}. In particular, transition metal dichalcogenide (TMDC) semiconductors exhibit remarkable properties, combining strong excitonic resonances with high refractive indices. Layered vdW crystals, especially when nanostructured, enable unique optical functionalities, unifying the roles of resonator and emitter within a single material platform \cite{verre2019transition}. This allows enhanced light-matter interactions and quantum emission to arise not from coupling to external cavities, but through the self-hybridization of the material’s own optical transitions with engineered qBIC resonances \cite{Weber2023,sortino2024optically}. The compact footprint and spectral tunability of vdW metasurfaces address key limitations of conventional optical cavities, enabling robust light-matter interactions at subwavelength scales and opening to tailored photonic resonance in vertical heterostructures \cite{sortino2025atomic}. 

Here, we combine momentum-resolved hypespectral imaging and ultrafast spectroscopy techniques to unambiguously unveil the quantum coherence of exciton-polaritons in bulk WS$_2$ qBIC metasurfaces at room temperature. We reconstruct the polariton momentum dispersion, which exhibit strong anisotropy and a negative effective mass along the qBIC resonant direction. Using ultrafast pump-probe spectroscopy, we observe that the exciton-polariton incoherent dynamics is strongly governed by the qBIC coupling and detuning, determining the recombination pathways between the uncoupled exciton reservoir at higher momenta and the negative-mass lower polariton (LP) state.
We investigate the coherent polariton dynamics by pushing the temporal resolution of pump-probe to sub-15-fs and by applying multidimensional coherent spectroscopy (MDCS) \cite{li2023optical}, which allows to establish a correlation between emission and excitation energies, as well as to analyse the electronic coherence and isolate quantum recombination pathways from correlated optical resonances. MDCS provides both a measurement of the intrinsic exciton-polariton coherence time of $\sim$110 fs in maximally coupled metasurfaces, and a direct evidence of strong polariton interactions. We reveal the unconventional coherent coupling between the LP branch and the uncoupled exciton (X) as characteristic oscillations in the pump-probe and MDCS signals with a $\sim$45 fs period, matching the LP-X energy gap. This can be captured by a three-eigenstate problem \cite{greten2024strong}, as supported by our theoretical model \cite{timmer2023plasmon}, extending the conventional two-level picture of strong-coupling.

\bigskip
\noindent
\textbf{Photonic band structure of strongly coupled metasurfaces}
The metasurface unit cell is composed of an asymmetric two-rod geometry, where their difference in length ($\Delta L$) introduces the symmetry-breaking conditions to control the qBIC resonance \cite{koshelev2018asymmetric,Weber2023}.
The main design parameters are shown in Figure \ref{fig:fig1}a. For a perfectly symmetric case ($\Delta L$ = 0) we have a dark BIC state. By adjusting the nanorod longitudinal dimension ($\Delta L \neq 0$) the qBIC resonance appears with a Fano-like line shape in the optical transmission \cite{koshelev2018asymmetric} and can be finely controlled to reach high Q factor values, mainly limited by fabrication resolution \cite{biechteler2025fabrication}. 
Additionally, the qBIC resonance can be spectrally tuned by applying a scaling factor $S$ to all in-plane unit cell parameters such that $S = P/P_{0} = L/L_{0}=\Delta L/\Delta L_{0}=$ w$/$w$_{0}$, except for the height, which is limited by the exfoliated crystal thickness. 
We exfoliated sub-50 nm thick WS$_2$ layers and patterned them via top-down nanofabrication methods following Ref.\cite{Weber2023}.
The crystalline structure of TMDCs consists of covalently bonded three-atom-thick layers separated by a vdW gap (Figure \ref{fig:fig1}b). 
Combined with the presence of highly polarizable transition metals, this gives rise to high refractive index values and stable excitonic resonances at visible wavelengths \cite{Khurgin2022} (Figure \ref{fig:fig1}c). 
Figure \ref{fig:fig1}d shows an electron microscope image of a fabricated WS$_2$ metasurface. 
We tailored the metasurface resonance to span across the WS$_2$ exciton energy (Figure \ref{fig:fig1}e). The experimental reflectance of fabricated metasurfaces with varying scaling factors (Figure \ref{fig:fig1}f) exhibits an avoided crossing when the exciton and qBIC resonances overlap, together with the presence of an uncoupled excitonic resonance.

We investigated the angular reflectance of qBIC metasurfaces, where structural asymmetry gives rise to a pronounced anisotropic response. The qBIC mode itself is intrinsically anisotropic, because of the nanorods' longitudinal dipole resonance, corresponding to charge oscillations along their long ($x$) axis that couple strongly to light polarized parallel to this axis. Light polarized along the short axis couples only weakly. This anisotropic behaviour can be leveraged to form saddle points, a momentum dispersion landscape favourable for lasing and exciton condensation \cite{Nigro2023}.
Figure \ref{fig:fig1}g shows the numerical simulations of the angle-resolved reflectance spectra for a WS$_2$ metasurface in vacuum and without the substrate (see also Supplementary Note 1), shown along the directions off the central $\Gamma$ point (inset Figure \ref{fig:fig1}g), specifically $\Gamma \rightarrow X$ for positive angles ($\theta_x$), and $\Gamma \rightarrow Y$ for negative angles ($\theta_y$).
The polariton dispersion exhibits strong anisotropy where a negative-mass polariton branch along the $x$-direction transitions to a predominantly flat, positive-mass dispersion in the $y$-direction (see Supplementary Note 2).
We employ a momentum ($k$)-space resolved hyperspectral imaging technique to reconstruct the full angular dispersion of the fabricated optical metasurfaces \cite{Genco2022}.  
This method, employing a common-path birefringent interferometer, enables to selectively excite the qBIC resonance while capturing the Fourier plane image from the sample across different polarizations, in a single acquisition scan, and extract a dataset containing both spectral and angular information (see Methods and Supplementary Note 3).
Figure \ref{fig:fig1}h shows the hyperspectral reflectance profiles of WS$_2$ metasurfaces with excitation linearly polarized along the $x$-axis (s-polarized), depicted along the $\theta_x$ and $\theta_y$ directions for positive and negative angles, respectively (Figure S1a).
The hyperspectral images clearly reveal a pronounced anisotropy between the main symmetry axes. The polariton dispersion exhibits a transition from a predominantly flat and positive curvature along $\theta_y$, to a negative parabolic shape along $\theta_x$, in agreement with numerical results. The reduction of visibility of the upper polariton (UP) in the angular reflectance is related to the presence of grating modes at the diffraction cut-off, or Rayleigh anomalies, characterized by a linear angular dispersion (see also Supplementary Figure 1).

\begin{figure*}
	\centering
	\includegraphics[width=1\linewidth]{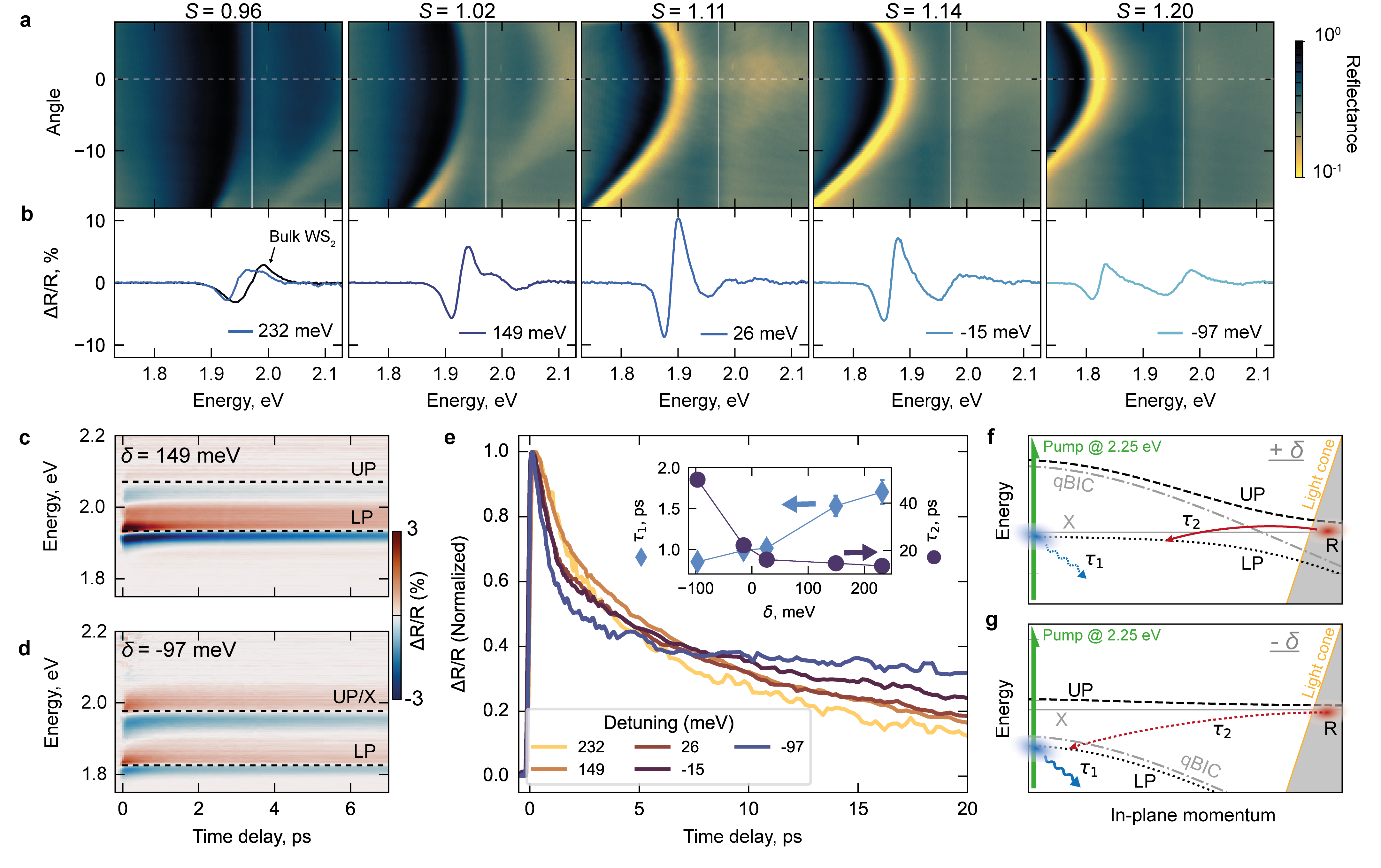}
\caption{\textbf{Ultrafast dynamics of exciton-polaritons in qBIC metasurfaces}
(a-b) Comparison of angle-resolved reflectance (a) and the zero-time delay transient reflectance spectra (b) for WS$_2$ metasurfaces with different detuning values. In the plot corresponding to a detuning of 232 meV, the black line represents the bulk reference WS$_2$ sample.
(c) Transient reflectivity map as a function of probe photon energy and pump-probe delay time for the sample with positive detuning (149 meV), showing a strong lower polariton (LP) resonance and a weak exciton (X) resonance.
(d) Corresponding map for negative detuning (-97 meV), where the upper polariton (UP) is nearly degenerate with the exciton.
(e) LP dynamics extracted from pump-probe measurements for samples with different detunings. Inset: fast and slow decay components obtained from fitting the LP dynamics as a function of qBIC detuning.
(f-g) Schematic illustration of LP relaxation mechanisms following high-energy excitation for positive (f) and negative (g) detuning, highlighting in the latter case the enhanced radiative decay, speeding up $\tau_1$, and the suppressed scattering from the exciton reservoir (R), slowing down $\tau_2$.}
	\label{fig:fig2}
\end{figure*}

\bigskip
\noindent
\textbf{Ultrafast dynamics of qBIC exciton-polaritons} 
To investigate the dynamics of the metasurface-coupled exciton-polaritons we initially performed collinear ultrafast transient reflection (TR) spectroscopy experiments. 
The sample was mounted in a custom microscopy setup capable of capturing both real-space and Fourier-space static reflectance images while simultaneously enabling TR spectroscopy with $\sim$ 150-fs time resolution (see Methods and Supplementary Note 4). 
This approach allows us to correlate the angle-resolved white light reflection (Figure \ref{fig:fig2}a) and ultrafast TR signals (Figure \ref{fig:fig2}b) for WS$_2$ metasurfaces at different detuning values. 
The exciton-qBIC detuning energy, $\delta$, is defined as $\delta = E_{qBIC} - E_X$, where $E_X$ and $E_{qBIC}$ are the exciton and qBIC resonance energies at $k_x=0$, respectively, being linearly dependent on the scaling factor, $S$, as shown in Figure S2b.

For TR experiments, we deliver ultrashort ($\sim$ 150 fs, after the objective) narrowband pump pulses (FWHM=40 meV, centered at 2.25 eV) and broadband (2.23-1.65 eV) probe pulses through an objective lens, focusing them on the sample with a spotsize of 3.5 $\mu$m, and collecting the back-reflected probe light. 
All the spectra in Figure \ref{fig:fig2}b, taken at zero-time delay, show almost symmetric positive-negative derivative-shaped signals around the energies of the exciton and polariton modes. 
For metasurfaces with large positive detunings, the qBIC resonance lies higher in energy than the exciton and mostly has a photonic character. The transient response is then dominated by the bulk uncoupled exciton (first panel in Figure \ref{fig:fig2}b, in black). As the qBIC is moved closer in resonance with the excitons, the TR signal begins to deviate from that of the bulk material, with the appearance of a strong LP response at lower energies. For negative detuning values, the exciton signal reappears, while the LP shifts to lower energies. 
Additionally, at a detuning value of 149 meV, we detect a weak signal at higher energies ($\sim$2.05 eV). By comparing its spectral position at zero-angle to the related angle-resolved spectrum in Figure \ref{fig:fig2}a, we ascribe this feature to the UP branch. 

Figures \ref{fig:fig2}c-d present the full TR maps for two selected metasurfaces (see also Supplementary Figure 6). For $\delta=$ 149 meV (Figure \ref{fig:fig2}c), the excitonic response is negligible compared to the signals from the polaritonic branches. Additionally, the UP signal decays much faster than the LP signal, suggesting a rapid scattering mechanism from higher-energy states to lower-energy ones \cite{virgili2011ultrafast}.
For negative detuning (Figure \ref{fig:fig2}d), the UP acquires a predominantly excitonic character, becoming nearly degenerate with the bare exciton, and both exhibit comparable lifetimes. While the TR spectral shapes resemble the bulk exciton ones, the time traces exhibit distinct relaxation behaviour depending on the detuning.

To understand the bare exciton relaxation dynamics, we analyse the exciton lineshape transient behaviour in bulk WS$_2$ by performing transfer matrix simulations of the dynamic response of the optical structure (see Supplementary Note 5). A prominent contribution of excitation-induced dephasing is observed at early times ($<$ 1 ps), causing an exciton linewidth broadening \cite{katsch2020exciton}. 
Our simulations take into account the excitation-induced modifications to both the real and imaginary part of the sample susceptibility (Fig.\ref{fig:fig1}c), although in TR experiments the former plays a major role.

We extract the LP dynamics for each detuning by tracking the temporal evolution of the reflectance dip around the LP energy in the TR maps (Figure \ref{fig:fig2}e). 
All traces exhibit a bi-exponential decay, characterized by a fast ($\tau_1\sim1$ ps) and a slow ($\tau_2\sim10$ ps) component. We fit the LP dynamics using a multi-exponential model with a rise and two decay terms (see Supplementary Note 6), with the extracted decay times shown in the inset of Figure \ref{fig:fig2}e. 
Similarly to uncoupled excitons in bulk WS$_2$, $\tau_1$ is associated with the rapid decay of the initial LP population. Starting from an exciton-like behaviour at positive detunings, $\tau_1$ decreases as the system moves towards negative detuning values. In contrast, $\tau_2$ exhibits the opposite trend, becoming progressively longer at negative detunings.
The opposite detuning dependence of $\tau_1$ and $\tau_2$ is a distinctive signature of polaritonic systems, consistent with earlier observations in planar quantum well microcavities \cite{sermage1996time}. It reflects the crossover between exciton- and photon-character of the polariton state: for positive detunings, the LP retains a strong excitonic character, leading to decay dynamics similar to uncoupled excitons. 

Conversely, at negative detunings, the LP becomes increasingly photonic, $\tau_1$ decreases while $\tau_2$ increases. 
In bulk TMDCs, $\tau_1$ has been attributed to electron and hole intervalley scattering processes \cite{nie2015ultrafast}. In strongly coupled systems, the polariton states act as an additional efficient radiative decay pathway \cite{shan2022brightening}, which can explain the decrease of $\tau_1$.
In our system, we attribute the variation in $\tau_2$ to the detuning-dependent exciton scattering from the dark exciton reservoir outside the light cone into the LP state \cite{fitzgerald2024circumventing}, mediated by the peculiar qBIC dispersion.
For positive detunings, the energy separation between the LP and the exciton reservoir at higher momenta becomes small (Figure \ref{fig:fig2}f), enabling an efficient scattering between the dark excitons and the LP, leading to a decrease of the slow LP decay time \cite{fitzgerald2024circumventing}. 
Instead, at negative detunings, the qBIC lies below the exciton energy over all momenta (Figure \ref{fig:fig2}g), inhibiting relaxation from the exciton reservoir. Phonon-assisted scattering, effective in planar microcavities for repopulating LP states \cite{tassone1997bottleneck}, is suppressed in this regime.
Supplementary Figure 8 shows a direct comparison between the exciton-like UP and photon-like LP dynamics in the negatively detuned sample, $S=1.20$, which further confirms the presence of a slow repopulation mechanism only for LP.

\begin{figure*}
	\centering
	\includegraphics[width=1\linewidth]{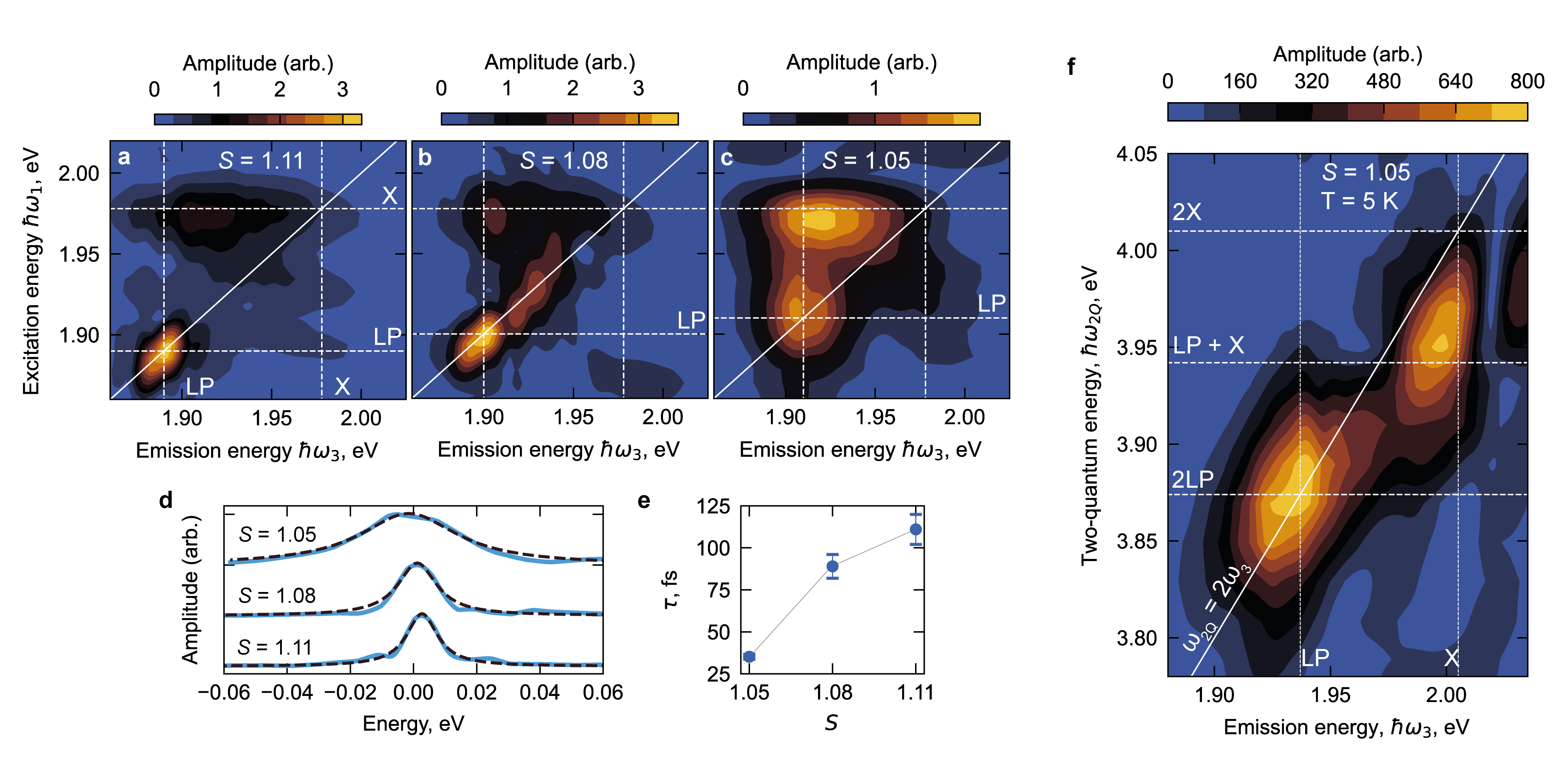}
	\caption{\textbf{1-quantum and 2-quantum MDCS of WS$_2$ qBIC metasurfaces}
(a-c) Amplitude maps of the 1-quantum box-CARS MDCS third-order nonlinear signal from WS$_2$ qBIC metasurfaces with different scaling factors ($S$), corresponding to increasing positive detunings for decreasing $S$.
(d) Fits of the diagonal peak for qBIC metasurfaces with varying $S$. Dashed lines represent Lorentzian fits to the data.
(e) Coherence times extracted from the full width at half maximum (FWHM) of the fits shown in panel (f).
(f) Two-dimensional 2-quantum spectrum of a WS$_2$ qBIC metasurface ($S = 1.05$) measured at 5 K. Peaks along the diagonal correspond to 2-quantum coherences involving identical states (either excitons or lower polaritons), while cross-peaks represent coherences involving two different transitions with distinct energies.}
	\label{fig:fig3}
\end{figure*}

\bigskip
\noindent
\textbf{Exciton-polariton coherence time} 
To disentangle the electronic correlations within our metasurface-coupled exciton-polaritons and their relevant recombination pathways, we employ MDCS \cite{Tollerud2017}, widely used to study interactions in quantum wells \cite{10.1126/science.1170274} and 2D semiconductors \cite{Moody2015, Muir2022}. MDCS enables to probe coupled quantum states in complex systems through energy-energy correlation maps. By using a sequence of three pulses to generate and manipulate quantum coherences and populations, MDCS captures both the amplitude and phase of the emitted signal as a function of excitation-emission energy and inter-pulse delays.
To fully capture the coherent population dynamics of our system, we employ two complementary MDCS approaches (see Methods for more details): the non-collinear boxCARS geometry \cite{Tollerud2017} and the partially-collinear pump-probe geometry \cite{fresch2023two}.

We begin our analysis from boxCARS MDCS experiments in a 1-quantum (1Q) rephasing experiment, where the coherence time (t$_1$) is scanned while keeping the population time (t$_2$) fixed.
This technique enables to disentangle homogeneous broadening from inhomogeneous broadening that arises from static disorder \cite{Tollerud2017}.
Using this approach, we studied the LP homogeneous broadening by analysing the diagonal peak in 1Q 2D spectra for metasurfaces close to zero detuning values (Figures \ref{fig:fig3}a-c). We observe that the amplitude signal exhibits a narrow waist along the anti-diagonal direction (Figure \ref{fig:fig3}a), indicating an inhomogeneously broadened transition, with a narrower homogeneous linewidth, hence longer dephasing time $T_2$, than expected from the estimate from linear spectroscopy \cite{Moody2015}. When increasing the detuning (Figure \ref{fig:fig3}b,c) the broadening of the LP peak indicates faster decoherence processes. 
By fitting the signal linewidth (Figure \ref{fig:fig3}d), we extracted values of the LP dephasing time up to $\sim$110 fs for the case of maximum coupling, and down to $\sim$35~fs for larger detuning (Figure \ref{fig:fig3}e).

Previous measurements at room temperature on monolayer WS$_2$ have shown dephasing times in the order of 50-75 fs \cite{wurdack2021motional,timmer2024ultrafast}, limited by exciton phonon scattering at low fluences, similar to those used here. At higher fluences, exciton-exciton interactions were shown to dominate. 
Hence, for large positive detunings, where the LP has large exciton fraction, the decoherence arises predominantly from interactions with phonons. 
In general, the exciton dephasing time has contribution from pure dephasing and energy relaxation. The reduction in decoherence observed under maximal coupling conditions is therefore ascribed to the mixing with the qBIC mode, which effectively mitigates phonon interactions.
Notably, for all the spectra we also observe a response at the (LP, X) cross peak, in coordinates ($\hbar\omega_3$, $\hbar\omega_1$), which becomes stronger relative to the (LP, LP) diagonal peak as the detuning increases. 
The appearance of this cross peak already indicates a coherent interaction between LP and X transitions.  As we reduce the $S$ factor (increased detuning) the exciton fraction of the LP is reduced, however the larger amplitude suggests that the coupling becomes stronger. This could be related to changes in the polariton bandstructure to a flat LP dispersion, introducing faster scattering from high-momenta exciton states as observed in previous TR experiments.

To help clarify the nature of the interactions between the X and LP transitions, we performed 2-quantum (2Q) coherent spectroscopy, with the same setup configuration, by changing the pulse ordering, having the first two pulses coincide, and scanning the delay of the final pulse \cite{Tollerud2017}. The 2Q signal is generated at twice the energy of the individual resonances, providing a background-free signal of the relevant electronic correlations in the system. In order to reduce noise and improve the coherent interactions, we performed the experiments in a closed-cycle helium cryostat at 5 K, and we selected the maximally coupled metasurface at this temperature taking into account the thermal shift of the excitons \cite{Weber2023}.
Figure \ref{fig:fig3}f shows the 2Q-spectrum as a function of the two-quantum energy ($\hbar\omega_{2Q}$) and the emission energy ($\hbar\omega_3$). The diagonal line runs on a 2:1 slope and a peak on this line represents the interaction between similar species. We observe the dominant presence of the (LP, 2LP) diagonal peak, confirming the strong polaritonic character. When looking at the cross-peaks, indicating the interaction involving species with different energy, the weight of the signal is strongly located at the (X, LP+X) cross peak. The pronounced cross-peak provides clear evidence of a correlated LP+X state, indicating mutual influence between the LP and the X transitions. This constitutes clear evidence of coherent interactions between the LP and X states, beyond what can be attributed to incoherent population transfer. Finally, the lack of a clear UP signal in MDCS experiments further supports its low oscillator strength and negligible weight in the system's dynamics.

   \begin{figure*}
       \centering
       \includegraphics[width=1\linewidth]{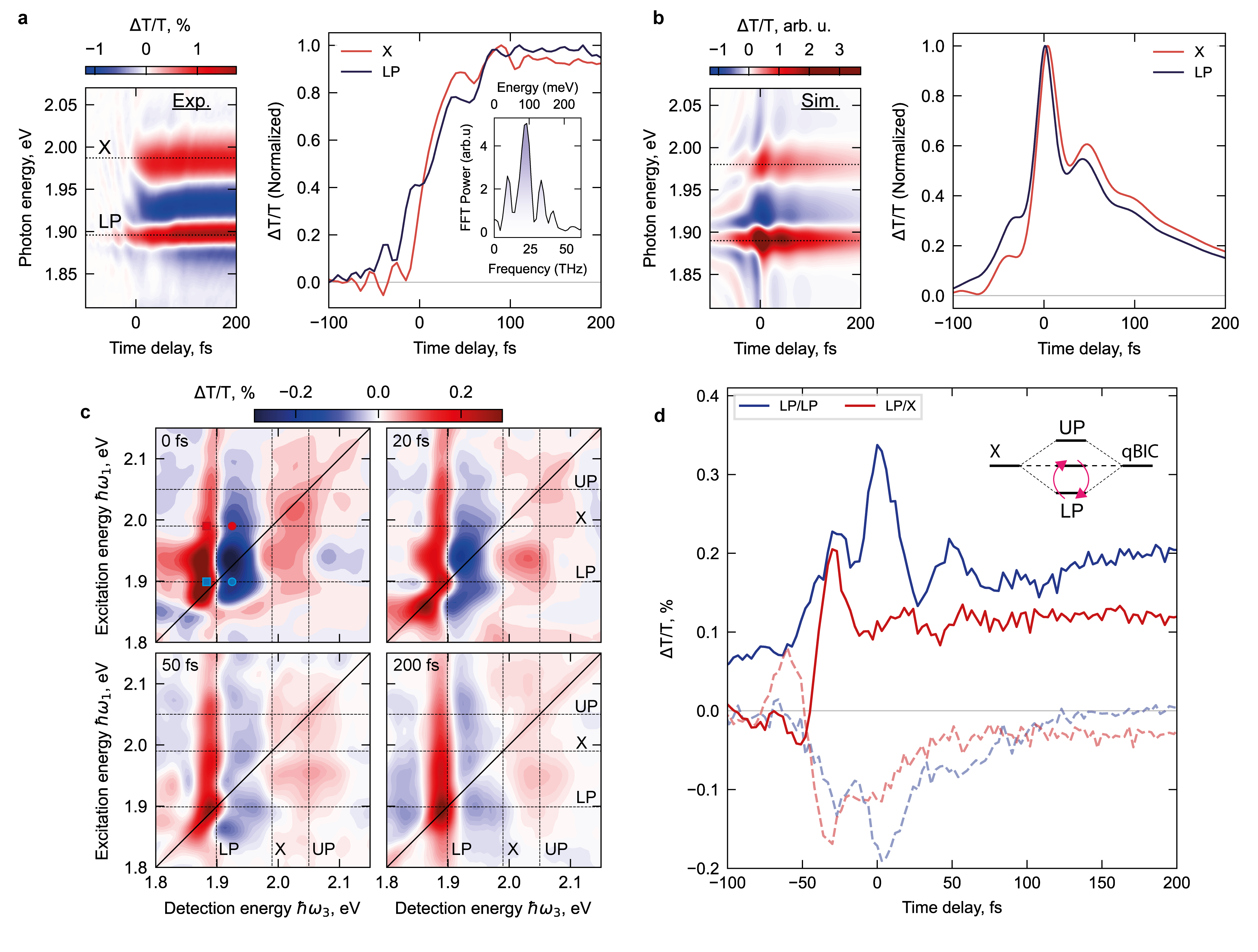}
       \caption{\textbf{Ultrafast coherent exciton-photon coupling} 
       (a) High temporal resolution ultrafast spectroscopy of a maximally coupled WS$_2$ metasurface ($S = 1.11$). Left panel: Differential transmission map. The dashed lines indicate the spectral positions of the lower polariton (LP) and the exciton (X). Right panel: Normalized transient transmission profiles extracted at the LP and X energies, respectively.
(b) Numerical modelling of the ultrafast response, as described in Supplementary Note 10. Right panel: Simulated differential transmission map. Left panel: Normalized differential transmission time traces at the LP and X energies, showing approximately three oscillation periods within the LP coherence time.
(c) Absorptive 2D maps of the WS$_2$ metasurface measured at different pump-probe delay times (t$_2$ = 0, 20, 40, and 150 fs).
(d) Ultrafast dynamics extracted from the 2D maps for the (LP, LP) diagonal peak and the (LP, X) cross-peak. The solid line indicates the positive signal from the transient response, the dashed lines the negative signal contribution, as depicted in the top left panel of (c) for square and circle marker, respectively. Inset: Energy levels of the strongly coupled system, with a coherent oscillations between the LP and X states (magenta arrows).
       }
       \label{fig:fig4}
   \end{figure*}
   
\bigskip
\noindent
\textbf{Ultrafast coherent exciton-polariton coupling}
To gain further insight into the exciton-polariton coupling we performed ultrafast pump-probe spectroscopy and MDCS experiments, increasing the temporal resolution by compressing laser pulses down to $\sim15$ fs (see Methods). 

Figure \ref{fig:fig4}a shows the differential transmission ($\Delta T/T$) response over a short window of 200 fs for a metasurface with $S$=1.11 ($\delta=26$ meV). A clear oscillatory behaviour at both the exciton and LP emission lines is observed in the $\Delta T/T$ map (Figure \ref{fig:fig4}a, left panel). When extracting the transient transmission profiles at the LP and X energies (Figure \ref{fig:fig4}a, right panel), we observe distinct oscillatory behaviours in the two curves, indicating coherent interactions lasting for 100-150 fs, which aligns closely with the previously measured coherence time. 
When probing the same sample with excitation polarization orthogonal to the nanorods (see Supplementary Note 8), suppressing the qBIC resonant coupling, the oscillations disappear confirming that the observed oscillatory behaviour is driven by the qBIC resonance. A Fourier transform analysis on the oscillations of the LP signal (Figure \ref{fig:fig4}a right Inset, see also Supplementary Note 9) reveals a prominent component at about 22.5 THz ($\sim$93 meV, period T$\sim$44 fs) ascribed to coherent Rabi oscillations between polariton and exciton states, in close agreement with the LP-X splitting measured in the static experiments. A less pronounced peak is also observed at $\sim$32.5 THz ($\sim$134 meV, period T$\sim$31 fs), linked to a larger splitting which could be attributed to the UP-LP coherent oscillations.

We modelled the ultrafast response based on the Lindblad master equation \cite{timmer2024ultrafast,timmer2023plasmon} (see Supplementary Note 10 for more details). Figure \ref{fig:fig4}b shows the computed results, which reproduce the observed oscillatory behaviour on temporal scales of approximately 45 fs. Here, the signal decays faster than the experimental results, due to the simplified three-level system model chosen in the simulations. This model does not take into account the effect of incoherent repopulation dynamics from dark excitonic states, responsible for the experimental long-lived transient response. 

To gain deeper insight into the coherent dynamics, we resolved LP-X Rabi oscillations by analysing diagonal and cross-peaks at different excitation energies in a broadband MDCS measurement with ultrashort pulses, covering $>$300 meV of bandwidth (Figure \ref{fig:fig4}c, see Methods).
Figure \ref{fig:fig4}d shows 2D maps as a function of the excitation and detection energy at four different t$_2$ time delays. 
The dominant features appear around detection energies corresponding to the LP, with the dispersive profile at early times indicative of excitation induced shift (EIS). Already at 0 fs, a structured signal landscape appears, with distinct cross-peaks at the excitation energies of LP, X and UP. By 200 fs, the excited population has significantly decayed, and the EIS correspondingly reduced, as evident by the dominance of the positive component that has moved closer to the LP energy. The peak modulation observed as a function of t$_2$ originates from interference between contributions from population pathways and from the LP-X coherent superposition, whose phase evolves with t$_2$.

We extracted the dynamics of the (LP, X) cross-peak and the LP diagonal peak as a function of the population time t$_2$ (Figure \ref{fig:fig4}e).
Notably, the signal from the LP diagonal peak exhibits pronounced oscillations with a period of $\sim45$ fs, directly reflecting coherent energy exchange between the LP and uncoupled X states. This timescale matches the LP-X energy splitting in our system and not the full Rabi splitting between LP and UP branches, emphasizing that the observed oscillations arise from polariton-exciton coupling rather than conventional LP-UP coherence. 
Focusing on the (LP, X) cross-peak, we again resolve clear oscillations, albeit with a more strongly damped amplitude. Two full oscillation cycles are visible in the temporal trace, providing direct evidence of coherent coupling between the LP and a reservoir of excitons.
Similar beating phenomena have been reported in plasmonic systems \cite{timmer2023plasmon}, attributed to spatial inhomogeneity in the near-field intensity, which seeds coexisting populations of coupled and uncoupled excitons.
We argue that their observation here in a strongly coupled dielectric metasurface points at the prominent role of high-momentum dark excitons in the coherent coupling with the qBIC mode, and thus in driving coherent polariton dynamics. 
The ability to observe and isolate these coherent exciton-polariton oscillations reveals a new regime of light-matter dynamics in vdW metasurfaces, with implications for reservoir-engineered polariton control.

\bigskip
\noindent
\textbf{Conclusions and outlook}

\noindent
In summary, our study uncovers the fundamental dynamics of room-temperature metasurface-coupled exciton-polaritons and establishes qBIC metasurfaces as a versatile and scalable platform for tailored strong light-matter interactions in subwavelength devices. 
By directly imaging the polariton momentum dispersion and combining complementary ultrafast spectroscopy techniques, we extract detuning-dependent relaxation dynamics, quantify coherence times, and resolve coherent coupling through characteristic oscillations in the transient optical response. 
The LP-X cross peaks observed in MCDS and coherent Rabi oscillations between such states point towards coherent coupling between X and LP, and a three-eigenstate system (LP, X, UP), beyond the simplistic 2x2 polariton Hamiltonian, as also reported for plasmonic systems \cite{timmer2023plasmon,greten2024strong}. 
These observations suggest that the eigenstates originate from the coupling of the qBIC with at least two distinct, but energy-degenerate, momentum-bright and dark excitons. The coupling is mediated by the near- and far-field contributions of the nanostructure, such that the resulting X eigenstate represents a qBIC-mediated superposition of the underlying exciton subsets \cite{greten2024strong,Weber2023}.
Engineering qBIC, shaping the underlying photonic bandstructure, provides a powerful platform for exploring light-matter hybrids, and vdW materials offer unprecedented control over the material properties, towards heterostructures with tailored electronic and optical properties and tunable through composition, twist angle, and interlayer hybridization.
Ultimately, the formation of coherent exciton-polariton populations in solid-state systems at room temperature not only advances light-matter interaction in low-dimensional systems, but also unlocks scalable and energy-efficient photonic architectures. As polariton systems become increasingly accessible through the integration of vdW materials and metasurfaces \cite{zotev2025nanophotonics,ling2021all}, we anticipate a new generation of polaritonic platforms capable of harnessing quantum coherence, strong nonlinearities, and ultrafast dynamics for practical quantum and classical information processing.

\bigskip
\noindent
\textbf{Methods}

\noindent
\textbf{Sample fabrication and characterization} The WS$_2$ sample were exfoliated from commercially available crystals (HQ Graphene) and fabricated following the procedure detailed in Ref.~\cite{Weber2023}. Static optical transmission of the fabricated metasurfaces was acquired in a WITec spectroscopy setup.

\noindent
\textbf{Hyperspectral momentum-resolved imaging}
The hyperspectral microscopy setup used to perform the momentum-space reflectivity measurements consists of a modified Leica microscope, with custom illumination and detection arms (see also Supplementary Note 3). For illumination, a white tungsten lamp is coupled to a multimode fiber (core diameter: 100 $\mu$m), whose tip is imaged on the sample using a collimation lens and a 100x microscope objective (N.A.=0.75). The back reflection is sent to the detection arm using a 50:50 broadband beamsplitter. A Fourier lens in the detection path allows to image the back focal plane of the objective lens on the camera, while the Translating-Wedge-Based Identical Pulses eNcoding System (TWINS) interferometer is inserted between the tube lens and a CCD camera \cite{Genco2022}. By changing the delay between the two replicas of the Fourier space image generated by the TWINS interferometer, an interferogram is recorded for each pixel of the image, whose Fourier transform yields the reflectivity spectrum. The final data consists of a 3D datacube (hypercube) of the reflectance as a function of $\theta_x$, $\theta_y$ and photon energy (see also Figure S3). The sample reflectance is subtracted from that measured on the substrate. 

\noindent
\textbf{Pump-probe spectroscopy}
For pump-probe measurements, we use broadband ultrashort frequency-tunable visible pump and probe pulses. Our setups are powered by an amplified Ti:sapphire laser generating 100 fs pulses at 800 nm (1.55 eV) with 2 mJ energy and 2 kHz repetition rate. A fraction of the laser output is used to drive a non-collinear optical parametric amplifier (NOPA) pumped at 400 nm (3.1 eV) by the second harmonic of the laser, generating broadband or narrowband visible pump pulses \cite{cerullo1998sub}. The pump pulses are modulated by a mechanical chopper at 500 Hz. For the broadband probe pulses, a white-light continuum is generated by focusing the 800 nm laser output on a 1-mm-thick sapphire plate. The probe beam is sent to a mechanical delay line which controls the delay between pump and probe pulses. For TR microscopy experiments, the pump and probe pulses are then combined by a 1-mm-thick beam splitter and focused on the sample using an objective lens with 8 mm focal length (NA=0.3), with an estimated excitation angle of $\pm 5^\circ$. The spatial overlap of sample and collinear pump and probe beams is obtained by a three-axis translation stage coupled to a home-built imaging system consisting of a white LED, for illumination, and a CMOS camera. To measure the $\Delta\mathrm{R}/\mathrm{R}$ spectra, the probe beam reflected by the sample is collected by the objective lens and delivered, via an additional beam splitter, to a dispersive spectrometer with a CCD camera. A set of long-pass and short-pass filters is used to cut out the excitation light from the signal. This setup allows also to perform static k-space reflectivity measurements.
For the high temporal resolution transient measurements, we implemented a non-collinear differential transmission setup based on a broadband NOPA as pump, compressed to $\sim$ 15fs using a chirped mirror pair. The probe beam was perpendicular to the sample surface, while the angle between pump and probe beams was $< 5^\circ$. 

\noindent
\textbf{Multi-dimensional coherent spectroscopy}
MDCS is a heterodyne detected four wave mixing (FWM) technique retrieving the third-order nonlinear polarization. The first pulse sets the system into a coherent superposition of the ground and excited states (t$_1$). The second pulse converts the superposition into a population state (t$_2$), and the third pulse creates the 3rd-order coherence, which radiates the signal at t$_3$. MDCS yields energy-energy maps that correlate excitation and emission energies, providing a means to disentangle coupled quantum states and their dynamics, where both the amplitude and phase of the signal are measured as functions of the emission energy and the delays between three excitation pulses. 
We employ two approaches for MDCS. The MDCS boxCARS scheme offers high signal-to-noise ratio and typically captures different signal components, known as rephasing and non-rephasing, in separate measurements. These components reflect different ways in which quantum states evolve and interfere over time: rephasing pathways are sensitive to inhomogeneous broadening, while non-rephasing pathways provide complementary information related to homogeneous processes. To reconstruct the full response of the system, these two contributions must be combined, often requiring careful phase-referencing.
In contrast, the partially collinear pump-probe geometry, using the TWINS to achieve passive interferometric stability, allows simultaneous acquisition of rephasing and non-rephasing signals in a single shot. This enables direct measurement of the absorptive part of the 2D spectrum and provides high stability and temporal resolution, making it ideal for capturing ultrafast dynamics.

For the box-CARS geometry MDCS measurements, we utilise a Yb:KGW amplifier system (Light Conversion, Pharos) generating 60$\mu$J 240 fs pulses at 1030nm, at 125~kHz repetition rate to pump (using the third harmonic signal) a NOPA. This generates broadband pulses centred at 1.9 eV. The four beams are generated using a 2D grating, and the delays on each beam are individually controlled using a pulse shaper based on a 2D spatial light modulator. The pulse shaper is also used to compress each of the pulses using the MIIPS algorithm \cite{xu2006quantitative}, giving close to transform-limited pulse durations of 22~fs.
The sample is mounted such that the x-axis of the sample (as defined in Fig.\ref{fig:fig1}) is horizontal, and the polarization of each of the pulses is set to be parallel to the x-axis.
We employ a box-CARS geometry (Figure \ref{fig:fig3}a) in which the three excitation beams are arranged at the corners of a square, before being focused onto the sample using a 75 mm focal length lens, which gives each beam a distinct wavevector. The system emits a third-order signal in a background-free direction, described by the wavevector $k_S = -k_1 + k_2 + k_3$. 
To extract both the amplitude and phase of the signal, we introduce a local oscillator (LO) that interferes with the emitted signal, generating a spectral interferogram. From this, the complete complex response of the system is reconstructed. The resulting 2D spectra are obtained by recording the amplitude and phase of the signal as functions of the emission frequency and one of the pulse delays. 
The geometry is arranged such that the centre of the box is normal to the surface, and the angle to each of the beams is $<2\deg$. Based on the dispersion curves in Fig 1, the energy does not change much over this range of angles. The fluence for each of the beams was $<4~\mu$J~cm$^{-2}$ per pulse. The overlapping spot size was larger than the mesas containing the WS$_2$ qBIC structure, so to reduce scatter and noise levels, the signal path was spatially filtered at an image plane to select just the light coming from the qBIC region.
For the 1Q rephasing measurements the arrival time of the pulse with wavevector $k_1$ was varied such that it always arrived first, while the delay between $k_2$ and $k_3$ was kept fixed at 200 fs. For the 2Q measurements, the pulse ordering was changed such that the $k_1$ pulse arrives last, and the delay between $k_2$/$k_3$ and $k_1$ was varied. For the 2Q measurements, the sample was also placed in a closed-cycle cryostat (Montana Instruments) and cooled to 5~K. 

The TWINS-MDCS setup employs a partially collinear pump-probe geometry. A Ti:sapphire laser (Coherent Legend) generates 100 fs pulses at 800 nm with a 1 kHz repetition rate. A NOPA produces broadband pulses (1.8-2.1 eV, $\sim$15 fs after chirped mirror compression), covering the LP, UP, and exciton states. We split the broadband pulse into pump and probe beams. Using the TWINS system \cite{rehault2014two}, we divide the pump into two phase-locked pulses with a controllable coherence time delay (t$_1$). The sample is excited by these pulses, and after a waiting time (t$_2$, varied from -120 fs to 200 fs in 3 fs steps), the probe measures the non-linear signal via self-heterodyne detection. A spectrometer (Stresing Entwicklungsburo) resolves the probe light, yielding detection energies. We obtain excitation energies by Fourier transforming the probe signal over t$_1$. This setup ensures high resolution in both excitation and detection energies and excellent temporal resolution.
Both pump and probe are vertically polarized and aligned to the x-axis direction. The pump impinges on the sample normally while there is a small angle with the probe of $\sim5 \deg$. The pump power was adjusted to $\sim20 \mu$J/cm$^2$.

\bigskip
\noindent
\textbf{Acknowledgements}
This work was funded by the European Union (ERC, METANEXT, 101078018, EIC, NEHO, 101046329 and EIC Pathfinder Open programme, QUONDENSATE, 101130384). Views and opinions expressed are however those of the author(s) only and do not necessarily reflect those of the European Union, the European Research Council Executive Agency, or the European Innovation Council and SMEs Executive Agency (EISMEA). Neither the European Union nor the granting authority can be held responsible for them. This project was also funded by the Deutsche Forschungsgemeinschaft (DFG, German Research Foundation) under grant numbers EXC 2089/1-390776260 (Germany’s Excellence Strategy), TI 1063/1 (Emmy Noether Program), SFB 1372 “Magnetoreception and navigation in vertebrates” (project number 395940726), Li 580/16-1, the Bavarian program Solar Energies Go Hybrid (SolTech) and the Center for NanoScience (CeNS). C.L. thanks the Nieders{\"a}chsische Ministerium f{\"u}r Wissenschaft und Kultur for support through DyNano and the Wissenschaftsraum ElLiKo). S.A.M. acknowledges the Lee-Lucas Chair in Physics. J.A.D. and J.O.T. acknowledge funding from the Australian Research Council (project number DP210102050, and CE170100039). We also acknowledge financial support by the European Union’s NextGenerationEU Programme with the I-PHOQS Infrastructure [IR0000016, ID D2B8D520, CUP B53C22001750006] “Integrated Infrastructure Initiative in Photonic and Quantum Sciences”.

\bigskip
\noindent
\textbf{Author Contributions}
L.S., A.G., C.C. and M.G. performed the ultrafast experiments, with contribution from F.G.. T.W. fabricated the samples. C.C. and M.C. performed the k-space hyperspectral measurements under the supervision of C.M. and G.V.. D.T. and C.L. performed the theoretical simulations. J.O.T. and J.A.D. performed the boxCARS MDCS experiments. L.S., A.G., C.C., M.G., M.C., and J.O.T. analyzed the data. L.S. and A.G. conceived the experiments and prepared the manuscript with contributions from all the co-authors. S.D.C, C.L., J.A.D., S.A.M, A.T. and G.C. supervised the project.

\bigskip
\noindent
\textbf{Data Availability}

\noindent
The data that support the findings of this study are available at: https://doi.org/10.5281/zenodo.15681631

\bigskip
\noindent
\textbf{Conflict of interest}

\noindent
The authors declare no competing interests.

\end{document}